%% file: main.tex
\documentclass[10pt,journal,compsoc]{IEEEtran}

\usepackage{amsmath}
\usepackage{amssymb}
\usepackage{amsthm}

\usepackage{graphicx}
\usepackage{array}
\usepackage{enumitem}

\usepackage{tikz}
\usetikzlibrary{arrows.meta,positioning,shapes.geometric,calc,backgrounds,fit}
\usepackage{listings}
\usepackage{xcolor}
\usepackage{xspace}

\usepackage{algpseudocode}

\usepackage{etoolbox}

\newtoggle{blind}
\togglefalse{blind}

\usepackage{main}

\usepackage[nocompress]{cite}
\usepackage{url}
\usepackage[hidelinks]{hyperref}

\AtBeginDocument{%
  }

\begin{document}
\bstctlcite{IEEEexample:BSTcontrol}

\title{Compositional Loop Invariant Inference for Multiphase Programs via Structured Neurosymbolic Reasoning}

\iftoggle{blind}{%
  \author{Anonymous Author(s)}
}{%
  \author{Mingxiu~Wang, Jiawei~Wang, Shuangxiang~Kan,
    Angela~Huo, Yuekang~Li, Xi~Zheng, and Xiao~Cheng\textsuperscript{*}%
    \IEEEcompsocitemizethanks{%
      \IEEEcompsocthanksitem M. Wang, X. Zheng, and X. Cheng are with
      Macquarie University, Sydney, NSW, Australia.
      E-mail: mingxiu.wang@students.mq.edu.au, xiao.cheng@mq.edu.au, james.zheng@mq.edu.au.
      \IEEEcompsocthanksitem J. Wang and Y. Li are with the University of New South Wales,
      Sydney, Australia. E-mail: jiawei.wang6@unsw.edu.au, yuekang.li@unsw.edu.au.
      \IEEEcompsocthanksitem S. Kan is with Singapore Management University, Singapore. E-mail: shuangxiangkan@outlook.com.
      \IEEEcompsocthanksitem A. Huo is with the University of Technology Sydney,
      Sydney, Australia. E-mail: huan.huo@uts.edu.au.}%
    \thanks{\textsuperscript{*}Corresponding author: Xiao Cheng (e-mail: xiao.cheng@mq.edu.au).}%
  }
}

\markboth{Author manuscript}%
  {Compositional Loop Invariant Inference for Multiphase Programs}

\IEEEtitleabstractindextext{%
  \input{content/0.Abstract}
}

\maketitle
\IEEEdisplaynontitleabstractindextext
\IEEEpeerreviewmaketitle

\input{content/1.intro_xiao}
\input{content/2.Background}

\input{content/3.moti_xiao}
\input{content/4.Method}

\input{content/5.Evaluation}
\input{content/6.Discussion}

\input{content/7.RelatedWork.tex}
\input{content/8.Conclusion}

\section*{Acknowledgments}
ChatGPT and Claude are used for language polishing. All scientific content is solely the authors' own work.

\bibliographystyle{IEEEtran}
\bibliography{main}

\end{document}

%% file: content/0.Abstract.tex
\begin{abstract}
Loop invariants are essential for deductive program verification, yet automatically inferring them for {multiphase programs}, whose loops contain conditional branches that induce multiple execution phases with distinct behaviors, remains an open problem. Existing LLM-based methods either reason about loops as a whole or use explicit branch information to generate atomic assertions jointly across branches, without maintaining complete candidate sub-invariants for independent generation and targeted repair.
We present \thistool{}, a structured neurosymbolic reasoning framework for multiphase invariant inference built on two interlinked mechanisms. First, compositional inference: \thistool{} statically decomposes a loop into symbolic {loop branch sequences}, conjunctions of the loop guard and branch conditions that characterize distinct execution scenarios, and guides the LLM to generate candidate sub-invariants for each one independently, so that each LLM query focuses on a single execution scenario rather than the entire loop. Second, counterexample-guided localized refinement: when SMT verification rejects a composed disjunctive invariant, \thistool{} traces the counterexample back through the branch structure to identify which specific loop branch sequence caused the failure and repairs only that sub-invariant, preserving all correct components. Both mechanisms are unified by a shared symbolic structure derived from the loop's branch semantics, which guides what the LLM generates, how candidates are composed, and where refinement is directed.
We evaluate \thistool{} on 316 linear arithmetic, 50 nonlinear arithmetic, and 54 multiphase benchmarks. \thistool{} solves 35 of 54 multiphase benchmarks, outperforming the best LLM-based baseline (29) by 20.7\% and the best traditional method (22) by 59.1\%. On linear/nonlinear benchmarks, \thistool{} remains competitive, solving 355 of 366 programs overall.
\end{abstract}

%% file: content/1.intro_xiao.tex
\section{Introduction\label{sec:introduction}}

Loop invariants~\cite{furia2014loop, d2008survey} are central to deductive program verification~\cite{fetzer1988program}, enabling formal reasoning about loop correctness via Hoare logic~\cite{hoare1969axiomatic}. Large language models (LLMs) have recently emerged as a promising approach to automatic invariant inference, overcoming the scalability limitations of traditional symbolic methods~\cite{colon2003linear,hojjat2018eldarica,clarke2000counterexample,bogomolov2017counterexample,jhala2006practical,mcmillan2010lazy} and the generalization difficulties of earlier learning-based methods~\cite{garg2014ice,garg2016learning,nguyen2017counterexample,si2020code2inv}. Recent LLM-based approaches~\cite{wu2024llm,cao2025clause2inv,LIU2026103387,li2026loris} follow a \emph{guess-and-check} paradigm: the LLM guesses candidate invariants or clauses, an SMT solver checks them, and counterexample feedback guides the next round of guessing. These methods show good performance on standard linear/non-linear benchmarks.

However, real-world loops often contain conditional branches that cause different iterations to follow different execution paths, each updating variables in its own way. Such programs are called \emph{multiphase programs}~\cite{riley2022multi} and arise naturally whenever loop variables cross domain boundaries that activate different branches (e.g., a loop that increments a counter~$i$ and executes one branch while $i < n$ but switches to another once $i \geq n$, so that the invariant must capture both phases). Generating correct invariants for multiphase programs is a recognized challenge~\cite{nguyen2014using,gulwani2009control}: the correct invariant is inherently a disjunction of phase-specific invariants, one per execution phase, whose complexity grows combinatorially with the number of branches. Despite long-standing interest, multiphase programs expose a critical blind spot in current LLM-based loop invariant inference techniques.

Many LLM-based methods struggle on multiphase loops because they are \emph{blind to the structure of the guess space}: in a multiphase loop, the program's branch structure induces a natural decomposition of the guess space into separate dimensions, one per execution phase, and the correct invariant is a disjunction of phase-specific clauses, one drawn from each dimension. These methods ignore this structure and collapse the multi-dimensional guess space into a flat, unstructured one, prompting the LLM with the entire program and producing candidates without distinguishing which dimension each belongs to. The resulting clauses are then combined without branch awareness: \lamforinv{}~\cite{wu2024llm} combines clauses exclusively via conjunction, unable to represent disjunctive invariants, while \clausetoinv{}~\cite{cao2025clause2inv} enumerates both conjunctions and disjunctions, causing the candidate pool to grow exponentially with the number of phases. When verification fails, the counterexample feedback is exploited only coarsely at program level, unable to localize which phase dimension caused the failure. Even approaches that improve feedback quality, such as LORIS~\cite{li2026loris}, which detects local reasoning errors in the LLM's proof steps, still operate on the loop as a whole and cannot attribute failures to specific phases.

PAL2Inv~\cite{pal2inv} is a recent path-sensitive exception. It applies abstract interpretation to obtain initial invariants, prompts an LLM to generate atomic assertions for every enumerated loop path, and uses counterexample-guided clause combination to assemble disjunctions of path-level conjunctions. This design makes path information explicit, but its generation step still asks the LLM to reason about all paths in one request, and its search state consists of path-indexed pools of atomic clauses rather than multiple alternative, complete sub-invariants for each path. Refinement therefore enriches and recombines clause pools instead of maintaining competing global compositions in which one selected phase component can be replaced while all other choices remain fixed. Thus, path sensitivity alone does not eliminate the simultaneous-reasoning and candidate-composition burdens of multiphase invariant inference.

These failures point to two core challenges. First, the LLM must reason about all phases simultaneously, yet the joint reasoning complexity overwhelms its capacity and the resulting candidate space grows combinatorially. Second, when verification fails, whole-loop feedback is too coarse to identify which phase is responsible, wasting repair effort on phases that are already correct. 
Our key insight is that the branch structure of a multiphase loop enables two interlinked mechanisms that address both challenges. First, \emph{compositional inference}: the distinct branch-condition combinations decompose the invariant synthesis task so that the LLM generates sub-invariants for each combination independently, reducing per-step reasoning complexity from the full loop to a single execution scenario. This decomposition also naturally admits multiple candidates per combination, providing candidate diversity that prevents premature commitment. Second, \emph{counterexample-guided localized refinement}: because these combinations partition the state space, any counterexample can be uniquely mapped to the responsible combination, so the framework repairs only the failed phase's sub-invariant while preserving all other phases. 

We realize this architecture in \thistool{}, a framework that tightly integrates neural generation and symbolic verification through a shared symbolic structure derived from the program's branch semantics. \thistool{} statically analyzes the loop's control flow to extract \myterm{}s, distinct combinations of branch conditions that partition the loop's execution space. This symbolic structure guides LLM reasoning at every stage: it determines what the LLM generates (one sub-invariant per \myterm{}), how candidates are composed (root-to-leaf disjunctions), and where refinement is directed (the specific \myterm{} that caused the failure). Through this structure, the framework realizes both contributions. For compositional inference, each \myterm{} becomes a level in a search tree, and the LLM generates multiple candidate sub-invariants for each level independently, maintaining candidate diversity across levels. For localized refinement, when the solver rejects a composed invariant, \thistool{} traces the counterexample back through the \myterm{} partition to identify the responsible \myterm{} and restricts repair to that \myterm{} alone, preserving all other sub-invariants.

We implement \thistool{} and evaluate its performance on three benchmark datasets: 316 linear arithmetic benchmarks, 50 nonlinear arithmetic benchmarks, and 54 multiphase benchmarks. The first two categories are widely used in prior work and serve as standard evaluation sets for state-of-the-art invariant synthesis tools. 
The multiphase benchmarks are drawn from the CHC-COMP repository~\cite{ruemmer2025aeval}, which provides realistic, publicly available verification tasks for inductive invariant inference and safety checking, ensuring fair and representative evaluation~\cite{de2023competition}.
We compare \thistool{} against state-of-the-art LLM-based tools (\lamforinv{}~\cite{wu2024llm}, \clausetoinv{}~\cite{cao2025clause2inv}, LORIS~\cite{li2026loris}, and PAL2Inv~\cite{pal2inv}), representative learning-based approaches (Code2Inv~\cite{si2020code2inv}, CLN2INV~\cite{ryan2019cln2inv}, and LIPuS~\cite{yu2023lipus}), and traditional methods, including the multiphase invariant synthesizer ImplCheck~\cite{riley2022multi}, the CHC solver Eldarica~\cite{hojjat2018eldarica}, and the software model checkers CPAchecker~\cite{beyer2011cpachecker} and UAutomizer~\cite{heizmann2013ultimate}, demonstrating its effectiveness across both standard and challenging benchmark settings.

Our contributions are summarized as follows:
\begin{itemize}[leftmargin=*]

    \item We identify limitations in current LLM-based loop invariant inference methods~\cite{cao2025clause2inv, wu2024llm, li2026loris, pal2inv} when applied to multiphase programs: most use whole-loop generation and composition, while even path-sensitive methods do not retain complete local alternatives that support counterexample-guided replacement of one selected \myterm{}'s sub-invariant.

    \item We propose \thistool{}, a framework built on two interlinked mechanisms: compositional inference, which decomposes invariant synthesis along \myterm{}s so that the LLM reasons about one execution scenario at a time, and counterexample-guided localized refinement, which traces SMT counterexamples through the \myterm{} partition to identify and repair only the failed sub-invariant while preserving all correct components.

    \item We evaluate \thistool{} on 316 linear arithmetic, 50 nonlinear arithmetic, and 54 multiphase benchmarks, demonstrating competitive loop invariant synthesis performance on linear/non-linear benchmarks and substantial loop invariant synthesis improvements on multiphase programs over both LLM-based methods, learning-based baselines and traditional verification engines. 
\end{itemize}


%% file: content/2.Background.tex
\section{Background} \label{sec:background}
This section introduces the concept of loop invariants and the principles for verifying their correctness.
A \textit{loop invariant} is a logical assertion $I$ that characterizes a property holding throughout the execution of a loop.
Consider a program of the form $\{\, P \,\}\;\textbf{while } B \;\textbf{do } S\;\{\, Q \,\}$, where $P$ denotes the precondition, $B$ the loop guard, $S$ the loop body, and $Q$ the postcondition.
Proving the correctness of such a loop amounts to finding an invariant $I$ that satisfies the following proof rule:
\[
\frac{
   P \Rightarrow I
   \quad
   \{\, I \wedge B \,\}\; S \;\{\, I \,\}
   \quad
   (I \wedge \lnot B) \Rightarrow Q
}{
   \{\, P \,\}\;\;\textbf{while } B \;\textbf{ do } S\;\;\{\, Q \,\}
}
\]
The three premises correspond to three verification conditions.
\textit{Initiation} ($P \Rightarrow I$) requires the invariant to hold upon loop entry.
\textit{Preservation} ($\{ I \wedge B \} \; S \; \{ I \}$) requires each iteration to preserve the invariant.
\textit{Termination} ($(I \wedge \lnot B) \Rightarrow Q$) requires the invariant, together with loop termination, to establish the postcondition.
While checking these conditions is decidable for many theories via SMT solvers, synthesizing a suitable invariant $I$ remains a key challenge, particularly for loops whose bodies contain conditional branches that give rise to distinct execution phases.

%% file: content/3.moti_xiao.tex
\section{Motivating Example} \label{sec:motivating_example}

In this section, we present a motivating example to illustrate the limitations of current LLM-based loop invariant generation methods and how our framework addresses these challenges.

\input{figure/motivating-example}

\autoref{lst:example} presents a C program containing a multiphase loop. The loop body exhibits three distinct execution scenarios, each characterized by a specific combination of the loop guard and branch conditions, highlighted with colored backgrounds. We call each such combination a \emph{\myterm{}} (formally defined in Definition~\ref{def:lbs}) and note that a single \myterm{} may span multiple program phases. This loop has three \myterm{}s: $unknown() \land x < z$, $unknown() \land x \ge z$, and $\neg unknown()$ (exit).

In the \myterm $unknown() \land x < z$, a loop invariant is relatively easy to identify:
\(
x < z \;\wedge\; z > 0 \;\wedge\; w \ge 0.
\)
However, in the \myterm $unknown() \land x \ge z$, the invariant becomes more complex due to two distinct entry conditions, namely $z > 0$ and $z \le 0$.
When $z > 0$, a valid invariant is:
\(
x \ge z \;\wedge\; y > z \;\wedge\; w + 2x = 3z.
\)
When $z \le 0$, the corresponding invariant is:
\(
x \ge z \;\wedge\; y > z \;\wedge\; w + 2x = 0.
\)

Consequently, the overall loop invariant must account for both phases, yielding the following disjunctive invariant:
\(
(x < z \;\wedge\; z > 0 \;\wedge\; w \ge 0)
\;\;\lor\;\;
\bigl(x \ge z \;\wedge\; y > z \;\wedge\;
\bigl((z > 0 \;\wedge\; w + 2x = 3z) \;\lor\; (z \le 0 \;\wedge\; w + 2x = 0)\bigr)\bigr).
\)

\noindent\textbf{Existing efforts and limitations.}
We illustrate the limitations of existing LLM-based methods by tracing how four representative tools handle this example.

\toola{}~\cite{wu2024llm} generates candidate invariants for the entire program, decomposes them into clauses such as $w \ge 0$, $x \ge 0$, $y > z$, and weak disjunctive terms like $(x = 0 \;\land\; w = 0) \;\lor\; (x > 0)$, and verifies each for Initiation and Preservation. Verified clauses are collected and used to check Termination. Because all clauses are generated at the program level, the tool cannot capture phase-specific properties and relies on global terms that are too coarse-grained for branch-specific relationships. \toola{} therefore fails on this multiphase program.

\toolb{}~\cite{cao2025clause2inv} also generates clauses from the full program and combines them via conjunction ($\land$) and disjunction ($\lor$), using counterexamples to prune unsatisfied candidates. However, this pruning can discard partially correct candidates: a clause such as $(x < z \;\land\; z > 0 \;\land\; w \ge 0)$, valid within the $x < z$ phase, may be eliminated because it does not satisfy the counterexample globally. \toolb{} therefore does not succeed on this example.

\toolc{}~\cite{li2026loris} refines a single candidate using feedback on the LLM's own reasoning: failed verification conditions are translated into first-order implications and checked by an SMT solver, with invalid implications fed back as hints. On this example, a monolithic guess such as $w = x$ is preserved in $\phi_1$ but violated in $\phi_2$, and the feedback correctly reports the failure. However, this flags only a local error within one global candidate and never suggests partitioning the state into $(x \le z \land w = x) \lor (x \ge z \land w = 3z - 2x)$. Because a missing disjunct is a structural rather than a local error, \toolc{} fails on this example.

PAL2Inv~\cite{pal2inv} explicitly enumerates loop paths and associates atomic assertions with each path, but two difficulties remain. First, it generates assertions for all paths in one response and excludes logical connectives from each assertion, leaving the symbolic combinator to recover the required structure. For the $x \ge z$ \myterm{}, the combinator must pair each equality with its corresponding guard and construct
\(
(z > 0 \land w + 2x = 3z) \lor (z \le 0 \land w + 2x = 0).
\)
Path labels distinguish $x < z$ from $x \ge z$, but do not determine this internal grouping: disjoining the equalities without guards admits spurious states, while conjoining the guarded alternatives is unsatisfiable. Second, a failed global candidate triggers further clause generation or recombination, without localizing the failure to one \myterm{} for replacement of its complete sub-invariant. A failure involving $x \ge z$ should direct repair to its guarded equalities while preserving the selected $x < z$ candidate. PAL2Inv can express the required invariant in principle, but path-indexed atomic assertions alone do not provide independently replaceable, complete local candidates, leaving both their logical composition and targeted repair to be resolved.

\noindent\textbf{Our solution.}\label{sec:oursolution}
Our framework tackles this example by first decomposing the loop into three \myterm{}s based on its branch structure: $\neg unknown()$ (exit), $unknown() \land x < z$, and $unknown() \land x \geq z$.
The LLM is then prompted to generate candidates for each \myterm{} \emph{independently}.
This scoped prompting focuses the LLM on one execution phase at a time: when reasoning about the $x \ge z$ \myterm{}, it only needs to consider how $w$ and $x$ evolve under $x \ge z$, rather than simultaneously reasoning about all branches.

However, committing to a single candidate per \myterm{} would be fragile: the correctness of each candidate depends on its combination with candidates from other \myterm{}s, and a wrong early choice has no fallback.
We therefore generate \emph{multiple} candidates per \myterm{} and organize them as a search tree, where each \myterm{} corresponds to a level and each root-to-leaf path assembles one complete disjunctive invariant.
For example, for the exit \myterm{} the LLM might produce $x \geq z \land w \leq 0$ and $w + x \leq z$, while for $unknown() \land x \geq z$, it might produce $x \geq z \land y > z \land w = -2(x - z)$ and $x \geq z \land y > z \land w + 2x \leq 3z$.

Consider the path that selects:
\begin{align*}
    &\text{exit} \to x \geq z \land w \leq 0, \\
    &\mathit{unknown}() \land x < z \to y > z \land x \leq z \land w = x, \\
    &\mathit{unknown}() \land x \geq z \to x \geq z \land y > z \land w = -2(x - z).
\end{align*}
Their disjunction:
\(
(x \geq z \land w \leq 0)
\;\lor\;
(y > z \land x \leq z \land w = x)
\;\lor\;
(x \geq z \land y > z \land w = -2(x - z))
\)
is sent to the SMT solver, which returns counterexample $(y = 4, z = 3, x = 3, w = 3)$.
Unlike the methods above, our framework does not discard everything.
Instead, it performs counterexample-guided localized refinement. Since the \myterm{}s partition the state space, the counterexample $(x = 3, z = 3)$ uniquely maps to the $unknown() \land x \ge z$ \myterm{}, and evaluating $\subinv_3$ under this valuation reveals that $w = -2(x - z)$ is the specific violated clause ($w = 3 \ne 0$).
We then refine \emph{only} this \myterm{}'s candidate, while preserving the already-correct candidates for exit and $x < z$. This phase-targeted repair is infeasible for prior methods, which operate at the whole-loop level and cannot attribute failures to specific \myterm{}s.
After four refinement iterations, the corrected invariant for this \myterm{} becomes:
\[
y > z \land x \geq z \land w = 3z - 2x \land x \leq y + z + 1.
\]

Combining all candidates yields the final invariant:
\(
(x \geq z \land w \leq 0)
\;\lor\;
(y > z \land x \leq z \land w = x)
\;\lor\;
(y > z \land x \leq z \land w = x \land x \leq y + z + 1)
\;\lor\;
(y > z \land x \geq z \land w = 3z - 2x \land x \leq y + z + 1).
\)
Although this invariant differs from the manually derived one and contains a redundant disjunct (the third is subsumed by the second), it is valid: it correctly captures the necessary conditions across all execution paths. Had the first tree path failed entirely, the framework could have backtracked to an alternative candidate (e.g., $w + 2x \leq 3z$) rather than restarting the whole process.

%% file: figure/motivating-example.tex
\begin{figure}[t]
\begin{lstlisting}[
    moredelim={**[is][\colorbox{blue!10}]{@b@}{@/b@}},
    moredelim={**[is][\colorbox{red!10}]{@r@}{@/r@}},
    moredelim={**[is][\colorbox{green!10}]{@g@}{@/g@}},
    escapechar=|,
]
int main() {
    int x, y, w, z;
    x = 0;  w = 0;
    assume(y > z);

    while (unknown()) {
@b@        if (x < z) {@/b@    |\color{blue!70!black}\sffamily\scriptsize // $\phi_1$: unknown() $\land$ x $<$ z|
@b@            w = w + 1;@/b@
@b@            x = x + 1;@/b@
@r@        } else {@/r@         |\color{red!70!black}\sffamily\scriptsize // $\phi_2$: unknown() $\land$ x $\geq$ z|
@r@            w = w - 2;@/r@
@r@            x = x + 1;@/r@
        }
    }
@g@    if (x > y + z) {@/g@    |\color{green!50!black}\sffamily\scriptsize // $\phi_3$: $\neg$unknown() (exit)|
@g@        assert(w <= 0);@/g@
    }
}
\end{lstlisting}
\caption{A multiphase loop with three \myterms{}. Each \myterm{} $\phi_i$ corresponds to a distinct combination of the loop guard and branch conditions, highlighted in \textcolor{blue!70!black}{blue} ($\phi_1$), \textcolor{red!70!black}{red} ($\phi_2$), and \textcolor{green!50!black}{green} ($\phi_3$).}
\label{lst:example}
\label{fig:multiphase}
\end{figure}

%% file: content/4.Method.tex
\section{Approach} \label{sec:approach}

We first introduce the concept of a \emph{loop branch sequence}, which is the key abstraction underlying our compositional invariant synthesis approach. We then present an overview of the \thistool{} framework, followed by detailed descriptions of its components.

\begin{definition}[\textbf{Loop Branch Sequence}]
\label{def:lbs}
Let a loop have guard $B$ whose body contains branch predicates $\mathcal{P}=\{p_1,\dots,p_n\}$.
Each maximal control-flow path through the loop body evaluates a subset of $k$ predicates $\{p_{i_1},\dots,p_{i_k}\}\subseteq\mathcal{P}$, where $1\le i_1<\cdots<i_k\le n$ are the indices of the predicates encountered along that path.
A \myterm{} is the conjunction of $B$ with the (possibly negated) predicates along such a path:
$\phi \;\triangleq\; B \land l_{i_1} \land \cdots \land l_{i_k}$,
where each $l_{i_j}\in\{p_{i_j},\neg p_{i_j}\}$ for $j=1,\dots,k$.
The formula $\neg B$ defines the \emph{exit case}.
The set $\Phi$ of all \myterm{}s together with $\neg B$ partitions the loop's execution space: for any state, exactly one element of $\Phi$ holds.
\end{definition}

\begin{example}
Consider a loop \texttt{while~(B) \{ if~(P) \{ if~(Q) S1; else S2; \} else \{ S3; \} S4; \}}.
Here $\mathcal{P}=\{P,Q\}$, yielding four \myterm{}s:
$\phi_1 = B \land P \land Q$,\;
$\phi_2 = B \land P \land \neg Q$,\;
$\phi_3 = B \land \neg P$,\;
$\phi_4 = \neg B$.
Note that $\phi_3$ does not involve $Q$, since the path through $\neg P$ never evaluates it.
\end{example}

\subsection{Overview}

\input{figure/framework-overview}

\autoref{fig:overview} presents the overall workflow of \thistool{}, which consists of four steps:

\textbf{Step~1: Loop Branch Sequence Extraction (\S\ref{sec:extraction}).}
We parse the input program to construct its control-flow graph (CFG), from which we identify loop structures and extract \myterms{}.
These \myterms{} decompose the loop's execution space into disjoint branches, enabling subsequent steps to reason about each branch independently rather than treating the loop as a monolithic whole.

\textbf{Step~2: Compositional Loop Invariant Synthesis (\S\ref{sec:synthesis}).}
Rather than asking the LLM to produce a complete invariant in one step, we perform compositional inference: each \myterm{} becomes a level in a symbolic search tree, and the LLM generates multiple candidate sub-invariants for each level independently, so that each query focuses on a single execution scenario. The tree naturally maintains a diverse pool of partial solutions that enables the localized refinement in Step~4.

\textbf{Step~3: Loop Invariant Verification (\S\ref{sec:verification}).}
Complete invariant candidates are formed by composing root-to-leaf paths via disjunction ($\lor$) and verified with an SMT solver. If any candidate satisfies all verification conditions, it is returned as the final invariant.

\begin{figure}[t]
\begin{algorithmic}[1]
\Require Loop header $v_0$ with guard $B$, CFG $G$
\Ensure Set of \myterms{} $\Phi$
\State $\Phi \gets \varnothing$
\Procedure{Collect}{$v,\; \mathit{guards}$}
    \If{$v = v_0$}
        \State $\Phi \gets \Phi \cup \{\mathit{guards}\}$;\ \Return
    \ElsIf{$v$ is a nested loop header}
        \State \Call{Collect}{$v.\mathit{exit},\; \mathit{guards}$}
    \ElsIf{$v$ is a conditional with predicate $p$}
        \State \Call{Collect}{$v_\mathit{true},\; \mathit{guards} \cup \{p\}$}
        \State \Call{Collect}{$v_\mathit{false},\; \mathit{guards} \cup \{\neg p\}$}
    \Else
        \State \Call{Collect}{$v.\mathit{next},\; \mathit{guards}$}
    \EndIf
\EndProcedure
\State \Call{Collect}{$\mathit{body}(v_0),\; \{B\}$}
\State $\Phi \gets \Phi \cup \{\neg B\}$
\State \Return $\Phi$
\end{algorithmic}
\caption{DFS-based \myterm{} collection algorithm.}
\label{alg:compact-collect}
\end{figure}

\textbf{Step~4: Counterexample-Guided Localized Refinement (\S\ref{sec:refinement}).}
When verification fails, we trace the counterexample through the \myterm{} partition to identify the specific \myterm{} responsible for the failure, then refine only that \myterm{}'s sub-invariant while preserving the sub-invariants of all other \myterm{}s. Prior methods operate at the whole-loop level and cannot attribute failures to specific phases, so they lack this phase-targeted repair capability. This targeted refinement loop repeats until a valid invariant is found or the time budget is exhausted.

\subsection{\MyTerms Extraction\label{sec:extraction}}

Given the CFG of a program, we first identify loop structures and extract the loop guard $B$. We then enumerate all feasible \myterms{} via the algorithm in \autoref{alg:compact-collect}, which performs a depth-first traversal of the CFG starting from the loop body entry.

At each conditional node with predicate $p \in \mathcal{P}$, the traversal forks into two branches, appending $p$ or $\neg p$ to the current guard list respectively. When the traversal reaches a back-edge to the loop header $v_0$, the accumulated guards form a complete \myterm{} $\phi = B \land l_{i_1} \land \cdots \land l_{i_k}$ (Definition~\ref{def:lbs}). Nested loops are treated as atomic blocks by jumping directly to their exit nodes, since their internal branches do not contribute to the outer loop’s \myterms{}. Finally, the exit case $\neg B$ is added as a separate \myterm{} to ensure the invariant covers loop-skipping executions.

The resulting set $\Phi$ partitions the loop’s execution space, so that any program state, and in particular any counterexample returned during refinement, can be uniquely mapped to a specific \myterm{}.

\autoref{alg:tot-full} presents the complete algorithm for our compositional loop invariant synthesis and refinement. Steps~2--4 correspond to synthesis, verification, and refinement, respectively, each of which we describe below.

\subsection{Tree-Structured Compositional Loop Invariant Synthesis\label{sec:synthesis}}

\begin{figure}[t]
\begin{algorithmic}[1]
\Require \myterms{} $\Phi=\{\phi_1,\phi_2,\dots,\phi_D\}$, program $x$, beam width $k$, timeout $T$
\Ensure Valid loop invariant $I$ or \textsc{Timeout}

\State $Y \gets \{[\ ]\}$ \Comment{Partial paths}
\For{$i=1$ to $D$} \label{alg:gene_start} \Comment{\textbf{Step 2: Synthesis}}
    \State $P_i \gets \Call{Generate}{x,\phi_i}$ \Comment{$n$ candidates for $\phi_i$}
    \State $Y' \gets \Call{Expand}{Y,P_i}$ \Comment{Extend paths}
    \State $V \gets \Call{Score}{x,Y'}$ \Comment{LLM scores paths}
    \State $Y \gets \Call{TopK}{Y',V,k}$ \Comment{Beam: keep top-$k$}
\EndFor \label{alg:gene_end}
\State $I \gets \Call{Verify}{Y}$ \label{alg:check} \Comment{\textbf{Step 3: Verify}}
\If{$I \neq \textsc{None}$} \Return $I$
\EndIf
\While{$\lnot\;\textsc{Timeout}(T)$} \label{alg:refi_start} \Comment{\textbf{Step 4: Refine}}
    \State $y^\star \gets \Call{SelectBest}{Y}$ \label{line:select}
    \State $(\sigma, j, \mathit{cond}, \mathit{BC}) \gets \Call{Diagnose}{y^\star}$ \Comment{Localize CE}
    \If{$\Call{IsAmbiguous}{j,\mathit{cond}}$} \label{alg:ambi} \Comment{Cross-branch CE}
        \State $j \gets \Call{JudgeBranch}{x,\sigma,j}$
    \EndIf
    \State $r \gets \Call{Refine}{x,y^\star,\sigma,j,\mathit{cond},\mathit{BC}}$ \Comment{Refine $\subinv_j$}
    \State $Y \gets \Call{Update}{Y,y^\star,j,r}$ \label{alg:refi_end}
    \State $I \gets \Call{Verify}{Y}$
    \If{$I \neq \textsc{None}$} \Return $I$
    \EndIf
\EndWhile
\State \Return \textsc{Timeout}
\end{algorithmic}
\caption{Compositional loop invariant synthesis and refinement algorithm.}
\label{alg:tot-full}
\end{figure}









This section details the synthesis phase of the algorithm in \autoref{alg:tot-full} (Lines~\ref{alg:gene_start}--\ref{alg:gene_end}). Rather than generating a monolithic invariant in one step, we organize the synthesis as a tree-structured symbolic search that decomposes the task along the \myterms{} extracted in Step~1.

\textbf{Tree structure.}
Let $\Phi = \{\phi_1, \dots, \phi_D\}$ be the set of \myterms{}, where $D = |\Phi|$. We formulate invariant generation as a tree search of depth $D$: the root represents the empty sequence $y = ()$, and each level $i$ corresponds to $\phi_i$. A node at level $i$ stores a partial path $y = (\subinv_1, \dots, \subinv_i)$, where $\subinv_j$ denotes the sub-invariant selected for $\phi_j$ ($j \le i$). When multiple candidates exist at each level, we write $\subinv^k_j$ for the $k$-th candidate sub-invariant at level~$j$.
This tree encodes the \emph{combinatorial search over sub-invariant selections}: branching at level~$i$ represents alternative sub-invariant proposals for~$\phi_i$, not alternative execution paths in the program. A root-to-leaf path assembles one complete disjunctive candidate by selecting exactly one proposal per \myterm{}.


\textbf{Expansion and pruning.}
At level $i$, the LLM takes the program $x$ and $\phi_i$ as input and generates $n$ candidate sub-invariants $P_i = \{\subinv^1_i, \dots, \subinv^n_i\}$ (\textsc{Generate} in \autoref{alg:tot-full}). Each existing path in $Y$ is extended with every candidate in $P_i$, yielding an expanded set $Y'$. To control combinatorial explosion, we score each extended path $y' \in Y'$ via an LLM-based evaluation prompt (\textsc{Score}) that assesses four criteria:
whether the sub-invariants along the path are mutually consistent, whether each imposes meaningful rather than vacuous constraints, whether the new sub-invariant matches the semantics of its \myterm{}, and whether the accumulated path is likely to satisfy the proof obligations when composed via disjunction.
The LLM returns a score for each path, and only the top-$k$ paths are retained (\textsc{TopK}), effectively applying beam search to bound the tree width.

\textbf{Synthesis prompt design.}
Given $(x, \phi_i)$, the generation prompt instructs the LLM to produce sub-invariants that (1)~hold whenever $\phi_i$ is active (\textit{Initialization}), (2)~are preserved across loop iterations even when execution transitions to another \myterm{} (\textit{Preservation}), and (3)~together with $\neg B$ imply the postcondition (\textit{Termination}). Each candidate is a conjunction of predicates over program variables, where multiple conjunctive terms represent disjunctive cases when a single \myterm{} admits distinct entry states.

After $D$ levels, each root-to-leaf path $y = (\subinv_1, \dots, \subinv_D)$ encodes a complete candidate invariant $I = \bigvee_{j=1}^{D} \subinv_j$. Beam search ensures that only the $k$ most promising candidates survive at each level, keeping the search tractable while preserving the candidate diversity that the subsequent localized refinement (\S\ref{sec:refinement}) relies upon.

\subsection{Loop Invariant Verification and Refinement}

\subsubsection{Loop Invariant Verification\label{sec:verification}}
As shown in \autoref{alg:tot-full} (Line~\ref{alg:check}), each surviving path $y = (\subinv_1, \dots, \subinv_D) \in Y$ is composed into a global candidate $I = \bigvee_{j=1}^{D} \subinv_j$ and checked against the three standard invariant proof obligations: \textit{Initialization}, \textit{Preservation}, and \textit{Termination} using an SMT solver. This composition is sound because each $\subinv_j$ is implicitly guarded by its corresponding \myterm{} $\phi_j$. Since the \myterms{} form a partition of the state space, the disjunction $\bigvee_j \subinv_j$ represents a case split over mutually exclusive regions, rather than requiring each sub-invariant to hold globally.

Specifically, the solver checks each candidate $I$ against three conditions.
\textit{Initiation} (\textbf{Init}: $P \Rightarrow I$) requires the precondition to imply that at least one sub-invariant holds at loop entry.
\textit{Preservation} (\textbf{Pres}: $\{I \wedge B\}\; S \;\{I\}$) requires that if any $\subinv_j$ holds before an iteration, then some $\subinv_{j'}$ (possibly $j' \neq j$) holds afterward. This cross-branch condition is critical for multiphase loops, where a single iteration may transition from one \myterm{} to another.
\textit{Termination} (\textbf{Term}: $(I \wedge \lnot B) \Rightarrow Q$) requires that upon loop exit, the disjunction of sub-invariants together with the negated guard implies the postcondition.
If any candidate satisfies all three conditions, it is returned as the valid invariant. Otherwise, for each failed candidate, the solver returns a counterexample $\sigma$ and the violated condition, which are passed to the refinement phase.

\subsubsection{Loop Invariant Refinement\label{sec:refinement}}
This section details the refinement phase of the algorithm in \autoref{alg:tot-full} (Lines~\ref{alg:refi_start}--\ref{alg:refi_end}). While the synthesis phase realizes compositional inference by decomposing generation along \myterm{}s, the refinement phase introduces the complementary mechanism of counterexample-guided localized refinement: it traces verification failures to a single \myterm{} and repairs only that phase's sub-invariant while preserving those of all other phases. This phase-level repair capability is unavailable to whole-loop methods. Together, compositional inference and localized refinement are the two domain-specific contributions of \thistool{}. The ablation study in \autoref{subsec:RQ3} confirms that removing localization causes a 23\% drop in solved benchmarks (from 35 to 27).

\textbf{Counterexample-guided localization.}
In each iteration, the most promising candidate $y^\star = (\subinv_1, \dots, \subinv_D)$ is selected (see \textsc{SelectBest} below). The \textsc{Diagnose} procedure invokes the SMT solver and returns a counterexample $\sigma$ (a concrete variable valuation), the violated Hoare condition $\mathit{cond} \in \{\textbf{Init}, \textbf{Pres}, \textbf{Term}\}$, and a set of broken clauses $\mathit{BC}$. Because $\Phi$ partitions the state space (Definition~\ref{def:lbs}), evaluating $\sigma$ against the \myterm{} guards uniquely identifies the responsible \myterm{} $\phi_j$. In the tree view, $\phi_j$ corresponds to level $j$ of the path $y^\star$, and $\mathit{BC}$ pinpoints which conjunctive predicates in $\subinv_j$ are violated under $\sigma$, providing fine-grained guidance for the LLM.

\begin{example}
Returning to the motivating example (\autoref{sec:motivating_example}), suppose the synthesis phase produces a path with
$\subinv_1 = (x \geq z \land w \leq 0)$ for the exit \myterm{},
$\subinv_2 = (y > z \land x \leq z \land w = x)$ for $unknown() \land x < z$, and
$\subinv_3 = (x \geq z \land y > z \land w = -2(x - z))$ for $unknown() \land x \geq z$.
The composed candidate $I = \subinv_1 \lor \subinv_2 \lor \subinv_3$ fails \textbf{Term} with counterexample $\sigma = (y{=}4,\, z{=}3,\, x{=}3,\, w{=}3)$.
Since $\sigma$ satisfies $unknown() \land x \geq z$, the failure is localized to $\phi_3$.
Evaluating each atomic predicate of $\subinv_3$ under $\sigma$:
$x \geq z$ holds ($3 \geq 3$),
$y > z$ holds ($4 > 3$),
but $w = -2(x - z)$ fails ($3 \neq 0$).
Hence $\mathit{BC} = \{w = -2(x - z)\}$, giving the LLM a precise target for repair.
\end{example}

\textbf{Localized refinement.}
The LLM is prompted to regenerate only $\subinv_j$ (\textsc{Refine} in \autoref{alg:tot-full}), receiving $\sigma$, $\mathit{BC}$, and the program context as input. The updated sub-invariant $r$ replaces $\subinv_j$ in $y^\star$ while all other levels remain unchanged (\textsc{Update}). The modified candidate is then re-verified. This targeted, level-wise edit ensures that each refinement iteration corrects exactly one sub-invariant without disturbing the rest.

\textbf{Ambiguity resolution (\textsc{IsAmbiguous}, Line~\ref{alg:ambi}).}
For \textbf{Init} and \textbf{Term} failures, the counterexample $\sigma$ is a single state, so evaluating $\sigma$ against the \myterm{} guards directly identifies the responsible \myterm{} $\phi_j$.
For \textbf{Pres} failures, however, the counterexample is a pair $(\sigma_{\mathit{pre}}, \sigma_{\mathit{post}})$ representing the state before and after one loop iteration. If both states satisfy the same \myterm{} guard, that \myterm{} is selected. The ambiguity arises when they fall into different \myterms{}: $\sigma_{\mathit{pre}} \in \phi_a$ and $\sigma_{\mathit{post}} \in \phi_b$ with $a \neq b$, representing a cross-branch transition. In this case, the failure may originate from either side:
(1)~$\sigma_{\mathit{pre}}$ is \emph{spurious}, i.e., not actually reachable during execution, meaning $\subinv_a$ is too weak and admits infeasible states, so the fix is to strengthen $\subinv_a$.
(2)~$\sigma_{\mathit{pre}}$ is \emph{genuine}, i.e., the iteration legitimately transitions from $\phi_a$ to $\phi_b$, but $\subinv_b$ fails to hold afterward. In this case, the fix is to repair $\subinv_b$.
Since reachability is undecidable in general, we use the LLM as an approximate oracle via \textsc{JudgeBranch}: given the program, precondition, and $\sigma_{\mathit{pre}}$, the LLM traces concrete variable updates to judge whether $\sigma_{\mathit{pre}}$ is reachable from the initial state. This is feasible because SMT-produced counterexamples typically involve small integer values reachable within a few iterations. If reachable, we refine $\phi_b$, and otherwise $\phi_a$.
This judgment is approximate: the next verification round rejects an invalid refined invariant, while later LLM judgments and \textsc{SelectBest} may recover by choosing a different component or candidate. Recovery is not guaranteed, as discussed in \autoref{sec:discuss}.

\begin{example}
In the motivating example, suppose $\subinv_2$ for the \linebreak $unknown() \land x < z$ \myterm{} is $(y > z \land x \leq z \land w = x)$, and the solver returns a \textbf{Pres} counterexample with $\sigma_{\mathit{pre}} = (x{=}2,\, y{=}4,\, z{=}3,\, w{=}2)$ in $\phi_2$ ($x < z$) and $\sigma_{\mathit{post}} = (x{=}3,\, y{=}4,\, z{=}3,\, w{=}3)$ in $\phi_3$ ($unknown() \land x \geq z$). Since the pre- and post-states belong to different \myterms{}, the failure is ambiguous. Here $\sigma_{\mathit{pre}}$ is genuinely reachable (starting from $x{=}0, w{=}0$ and iterating twice with $x < z$), so the problem is that $\subinv_3$ does not accommodate states entering from $\phi_2$. The framework therefore refines $\subinv_3$, not $\subinv_2$.
\end{example}

\textbf{Candidate selection (\textsc{SelectBest}, Line~\ref{line:select}).}
When all candidates in $Y$ fail verification, a single candidate must be chosen for refinement. A na\"ive heuristic that ranks by the number of satisfied Hoare conditions tends to favor vacuously true invariants (e.g., a trivially true candidate can satisfy both \textbf{Init} and \textbf{Pres}), making count-based ranking unreliable. We instead adopt an LLM-based selection strategy: a prompt presents the program~$x$, all current candidates in~$Y$, and their corresponding counterexamples, and asks the LLM to select the candidate whose sub-invariants impose the most specific (non-vacuous) constraints, whose violated clauses are fewest and suggest a localized fix rather than a structural mismatch.
Because \textsc{SelectBest} is invoked at every refinement iteration (Line~\ref{line:select}), it naturally enables backtracking: if repeated refinement of one path does not converge, a different path from $Y$ may be selected in a subsequent iteration, effectively switching to an alternative candidate at the problematic level while reusing the sub-invariants at other levels.

%% file: figure/framework-overview.tex
\begin{figure*}[t]
\centering
\includegraphics[width=0.9\textwidth]{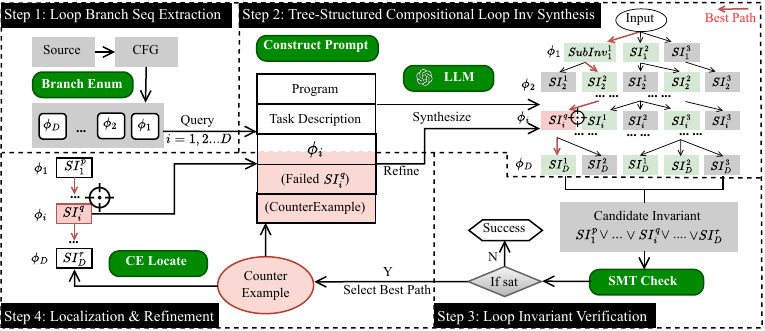}
\caption{Overview of the \thistool{} framework.}
\label{fig:overview}
\end{figure*}

%% file: content/5.Evaluation.tex
\section{Evaluation} \label{sec:evaluation}

In this section, we address the following research questions through an empirical evaluation of \thistool{}:
\begin{itemize}
    \item[\textbf{RQ1}] How effective is \thistool{} on multiphase loop benchmarks compared to state-of-the-art tools?
    \item[\textbf{RQ2}] How does \thistool{} generalize to linear and nonlinear benchmarks relative to existing approaches?
    \item[\textbf{RQ3}] What is the impact of individual components of \thistool{} on its overall performance?
    \item[\textbf{RQ4}] How do the effectiveness and runtime of \thistool{} vary as the number of loop-body branches increases?
\end{itemize}

\subsection{Setup}
\textbf{Benchmark suite.}
We evaluate \thistool{} on three benchmark suites comprising a total of 420 verification tasks: (1) 316 linear arithmetic benchmarks from \toola{}~\cite{wu2024llm}, (2) 50 nonlinear(NL) arithmetic benchmarks from the \toolb{}~\cite{cao2025clause2inv} benchmark suite, and (3) 54 multiphase arithmetic benchmarks from the CHC-COMP repository~\cite{ruemmer2025aeval}. Linear benchmarks contain loops whose invariants can be expressed as linear arithmetic constraints (e.g., $x + y \leq 10$), while nonlinear benchmarks require polynomial or more complex relationships (e.g., $x \times y \geq n$). These benchmarks are widely used in the loop invariant synthesis literature and cover a diverse range of program structures and invariant complexities.

\noindent\textbf{Baselines.}
We compare \thistool{} against a diverse set of tools spanning three paradigms:
(1)~LLM-based methods, including \toolb{}~\cite{cao2025clause2inv}, \toola{}~\cite{wu2024llm}, LORIS~\cite{li2026loris} and PAL2Inv~\cite{pal2inv};
(2)~learning-based methods, including Code2Inv~\cite{si2020code2inv} (reinforcement learning), CLN2INV~\cite{ryan2019cln2inv} (neural network-based), and LIPuS~\cite{yu2023lipus} (RL-based);
(3)~traditional methods, including the multiphase invariant synthesizer ImplCheck~\cite{riley2022multi}, the CHC solver Eldarica~\cite{hojjat2018eldarica}, and the software model checkers CPAchecker~\cite{beyer2011cpachecker} and UAutomizer~\cite{heizmann2013ultimate}. ImplCheck discovers phase guards through model-based projection and learns phase-specific lemmas, combining them as implication invariants. Eldarica produces invariants directly by solving constrained Horn clauses. CPAchecker and UAutomizer verify safety properties via unreachability checking. For these tools, we extract loop invariants from their correctness witnesses and validate them with our SMT-based checker against the three standard Hoare conditions.

\noindent\textbf{Evaluation metrics.}
We evaluate the performance of \thistool{} and the baseline tools using the following metrics:
\begin{itemize}[leftmargin=*, topsep=0pt, itemsep=0pt]
    \item \textbf{\# Solved:} The total number of benchmarks for which the tool successfully synthesizes a valid invariant within the given time limit.
    \item \textbf{Avg. Time (s):} The average time taken by the tool to solve each benchmark, measured in seconds.
    \item \textbf{Avg. Cost (\$):} The average monetary cost incurred by the tool for using the LLM API, measured in US dollars.
\end{itemize}

\noindent\textbf{Implementation.}
We implement \thistool{} in Python. The front-end uses \texttt{tree-sitter}~\cite{tree-sitter} to parse input C programs and construct ASTs, from which we derive the CFG and extract \myterms{}. Candidate invariants are checked with Z3~\cite{deMoura2008z3}.

\noindent\textbf{Experimental setup.}
All experiments were conducted on a 20-core Intel Core Ultra 7 265K CPU (up to 5.5~GHz) running Ubuntu 24.04 LTS. For LLM-based approaches, we use GPT-5.1. For the tree-structured search, we set the parameters as follows: the number of generated samples per step is 5, the number of evaluation samples is 3, and the beam width (i.e., number of selected candidates) is 5.
Each tool was allocated a maximum timeout of 600 seconds per benchmark, consistent with the budget used by \toola{}~\cite{wu2024llm} and \toolb{}~\cite{cao2025clause2inv}. We used the same version of the LLM API for all LLM-based tools to ensure a fair comparison. 

\subsection{RQ1: Multiphase Benchmarks} \label{subsec:RQ1}

\input{table/multiphase-results}


\autoref{tab:multiphase-results} summarizes performance on the 54 multiphase benchmarks. \thistool{} solves 35 tasks, compared with 25 for \toola{}, 20 for \toolb{}, and 18 for \toolc{}. Among these three LLM-based baselines, \toola{} has the highest coverage. \thistool{} solves 40\% more tasks. The strongest traditional method, ImplCheck, solves 22 tasks, followed by Eldarica with 20, giving \thistool{} a 59.1\% improvement over ImplCheck. Learning-based methods solve at most 5 tasks, indicating limited effectiveness on this suite. \thistool{} takes 167.48 seconds on average and costs \$0.226 per successful case: it is cheaper than \toola{} and close to \toolc{}, while \toolb{} is substantially cheaper.

PAL2Inv~\cite{pal2inv} is the strongest baseline on this suite, solving 29 tasks, six fewer than \thistool{}. The relative improvement is 20.7\%. Its path-sensitive design and initial invariants from abstract interpretation provide useful guidance for multiphase reasoning. It also has lower average runtime and API cost than \thistool{} (104.014 seconds and \$0.174). However, its refinement enriches and recombines path-indexed pools of atomic clauses rather than replacing one complete branch-specific sub-invariant while preserving the others. As these pools grow, the search must consider more clause combinations, and increasingly disjunctive candidates can burden SMT verification. This creates a potential scalability bottleneck. Abstract interpretation also supplies guidance whose precision depends on the chosen abstract domain. Thus, strong multiphase performance does not guarantee equally broad coverage across other program classes. PAL2Inv has the lowest linear and nonlinear coverage among the evaluated LLM-based methods, a contrast examined in \autoref{subsec:RQ2}.

Many of the remaining unsolved cases involve modulo ($\%$) operations in branch conditions, which introduce highly non-monotonic and cyclic execution patterns. These behaviors make multiphase programs particularly challenging, as execution repeatedly alternates between \myterms{} rather than following clearly separable phases. Consequently, it becomes difficult to capture program behavior using simple branch-based invariants. This pattern is common in multiphase benchmarks and poses a fundamental challenge for invariant synthesis, often requiring sophisticated reasoning even for human experts. A more detailed analysis of such cases is provided in \autoref{sec:case_study}.

\thistool{}'s higher coverage is consistent with its compositional design. Each \myterm{} defines an inference component for which the LLM generates several complete candidate sub-invariants. Tree search selects one candidate per component, and their disjunction forms a complete loop invariant. This separates local inference from global composition: the LLM reasons about the arithmetic relations within one execution case, while the search explores alternative combinations across cases. The same component structure supports refinement: a verification counterexample identifies the component to repair, whose selected sub-invariant is replaced while the choices for other components remain fixed. Local alternatives and targeted replacement therefore work together to retain useful partial solutions across verification rounds.

\subsection{RQ2: Linear/Nonlinear Benchmarks} \label{subsec:RQ2}

\input{table/linear-nonlinear-results}

As shown in \autoref{tab:existing-results}, \thistool{} solves 311 of the 316 linear benchmarks (98.4\%). This ties \toola{} for the highest linear coverage, while exceeding \toolb{} by 4 benchmarks and \toolc{} by 24. It also solves 19 more linear benchmarks than the strongest traditional baseline, UAutomizer, and 100 more than the strongest learning-based baseline, CLN2INV. The near-ceiling result shows that the branch-aware design does not sacrifice effectiveness on simpler control-flow structures. When a loop contains only one or a few \myterms{}, decomposition produces a correspondingly small set of focused inference tasks. The framework therefore retains the benefit of branch-level reasoning without creating a large compositional search space, which helps it remain effective on conventional linear benchmarks.

The advantage is more pronounced on the nonlinear suite. \thistool{} solves 44 of 50 benchmarks (88\%), the highest coverage among all evaluated tools. This is 1 more than \toolc{} (43), 6 more than \toolb{} (38), 13 more than \toola{} (31), and 15 more than UAutomizer (29). None of the evaluated learning-based methods solves a nonlinear benchmark. These tasks require invariants containing polynomial expressions or relationships among several variables. The zero results for the learning-based approaches show that this suite lies outside their effective coverage, while the strongest traditional method reaches only 29 tasks. By separating branch-specific obligations, \thistool{} presents the LLM with smaller inference problems and preserves a correct local relation when another branch requires refinement. The narrow one-benchmark margin over \toolc{} also indicates that improved local feedback is effective on nonlinear reasoning, whereas \thistool{}'s larger gains emerge when such feedback is combined with explicit branch decomposition.

PAL2Inv shows a different pattern from its strong multiphase result: it solves 227/316 linear tasks (71.8\%) and 15/50 nonlinear tasks (30.0\%), the lowest coverage among the evaluated LLM-based methods on both suites. \thistool{} solves 84 more linear and 29 more nonlinear tasks. This contrast is consistent with limits of PAL2Inv's dependence on guidance from abstract interpretation and its growing search over clause combinations. When the abstract domain provides only coarse information, the LLM must supply more of the missing relations. Adding atomic clauses then enlarges the combinations considered without identifying one complete local candidate to replace. \thistool{} instead bounds its retained candidate paths with beam search and uses localized refinement to replace a selected component.


Overall, \thistool{} achieves the highest multiphase coverage while generalizing well to the linear and nonlinear suites, solving 355 of 366 tasks. It ties for the best linear coverage (311/316) and achieves the highest nonlinear coverage (44/50). These results indicate that branch-aware compositional inference and localized refinement retain effectiveness on simpler control-flow structures while supporting more complex nonlinear reasoning.

\subsection{RQ3: Ablation Analysis} \label{subsec:RQ3}

\input{table/ablation-results}

We conduct two ablation studies on the multiphase benchmarks to evaluate the two domain-specific contributions of \thistool{}: compositional inference (decomposing generation along \myterm{}s) and counterexample-guided localized refinement.

The first variant, \thistool-NL (\textit{No-Localization}), removes the localization mechanism in the refinement process. Instead of tracing counterexamples to identify and refine the specific failing \myterm{}, the LLM is prompted to regenerate the entire invariant based on the program, the counterexample, and the current (incorrect) invariant. This directly tests the value of localized refinement. The second variant, \thistool-CoT, replaces the tree-structured search with a single Chain-of-Thought (CoT)~\cite{wei2022chain} generation process, which is equivalent to setting beam width $k = 1$, thereby removing structured exploration over \myterms{} while retaining localized refinement.

As shown in \autoref{tab:ablation-results}, both variants achieve lower success rates compared to the full \thistool{} framework. \thistool-NL solves 27 out of 54 benchmarks, a drop of 8 compared to the full framework. Without localization, the LLM regenerates the entire invariant upon receiving a counterexample, which often introduces errors into previously correct sub-invariants. A representative example of this regression behavior is discussed in \autoref{sec:case_study}.

The \thistool-CoT variant solves only 20 benchmarks, the lowest among all configurations. With $k = 1$, each \myterm{} is assigned a single candidate, leaving no fallback options for either synthesis or refinement. The 43\% drop in solved benchmarks (from 35 to 20) confirms that maintaining multiple candidates per \myterm{} provides substantial benefit, as the tree-structured search explores diverse candidates and selects the combination that best satisfies local verification.

The full \thistool{} framework incurs moderately higher runtime and cost, which is expected given that it solves 8--15 more benchmarks, many of which are inherently complex cases requiring multiple refinement iterations. These results confirm that the two domain-specific mechanisms address complementary failure modes: localized refinement prevents cross-branch regression during repair (\thistool-NL drops 8 benchmarks), while compositional inference with candidate diversity prevents single-point generation errors during initial synthesis (\thistool-CoT drops 15 benchmarks). Both are essential. Beam search provides the candidate pool but is itself a standard technique. The novelty lies in the compositional decomposition and the localized refinement it enables.

\subsection{RQ4: Effect of Branch Count} \label{subsec:RQ4}

To examine how branching affects \thistool{}, we stratify all 420 benchmarks by control-flow structure. The original linear, nonlinear, and multiphase suites follow categories established by prior work, but arithmetic and branching characteristics can overlap: a linear-arithmetic benchmark may still contain several conditional paths. Pooling the suites and grouping them by branch count allows us to examine branching independently of these suite labels.

We count the distinct non-exit loop-body branch sequences extracted from the CFG, excluding the exit case because it does not execute the loop body. A straight-line loop therefore has one branch sequence, while a loop with an internal conditional has two or more. This gives 176 single-branch and 244 multi-branch benchmarks. \thistool{} solves 174/176 single-branch tasks (98.9\%), three more than the strongest LLM-based baseline, and 216/244 multi-branch tasks (88.5\%), 13--76 more than the LLM-based baselines. Its success rate is 10.3 percentage points lower on the multi-branch subset, but its larger coverage margin over the baselines is consistent with decomposition being useful when inference must coordinate multiple execution cases.

\input{table/branch_count_result}

\autoref{tab:branch-count-results} provides a finer breakdown. Success rates vary from 85.6\% to 94.1\% across the groups with two or more branches and do not decrease monotonically with branch count. Median runtime, however, increases steadily from 19.3 seconds for one branch to 117.9 seconds for five or more. This trend is consistent with the framework's design: additional branches create more sub-invariant generation tasks and enlarge the compositional search, while each generation query focuses on one execution case and localized refinement targets one component. Thus, the observed runtime grows with branch count, while coverage remains high across the groups. Semantic difficulty, including nonlinear or periodic updates and relationships spanning branches or phases, can still make local candidates difficult to infer. Because the groups differ in size and program semantics, these aggregate results do not isolate the causal effect of branch count. Nevertheless, they show increasing runtime without a monotonic decline in success on this corpus.

\subsection{Case Study} \label{sec:case_study}
We analyze two representative cases. The first illustrates how compositional inference and localized refinement work together on a multiphase program. The second examines a periodic program beyond the multiphase setting, offering insights into future directions.

\subsubsection{Multiphase Invariant Synthesis}

\input{figure/multiphase-case-study}

The program in \autoref{lst:example-multi} exhibits clear multiphase behavior, where different ranges of $x$ correspond to distinct update patterns for $y$. \thistool{} addresses this by decomposing invariant synthesis along the four \myterms{}, so that each phase is reasoned about independently. An initial invariant generated by \thistool{} takes the form:
\(
(x < 5000 \land x \ge 4000 \land y = 4x - 16000)
\;\lor\;
(x < 4000 \land y = x)
\;\lor\;
(x \ge 6000 \land y + x = 0)
\;\lor\;
(x \ge 5000 \land x < 6000 \land y = -3x + 25000).
\)

We verify this candidate using an SMT solver, which returns a counterexample $(x = 5000, y = 4000)$. Through counterexample-guided localization, the failure is traced to the \myterm{} $(\texttt{unknown()} \land x \ge 5000 \land x < 6000)$, identifying the violated clause as $y = -3x + 25000$. Accordingly, only the sub-invariant for this \myterm{} is updated, while all other components are preserved.

By iteratively applying this targeted refinement process, \thistool{} converges to a correct invariant:
\(
(x < 5000 \;\land\; x \ge 4000 \;\land\; y = 4(x - 4000) + 4000)
\;\lor\;
(x < 4000 \;\land\; y = x \;\land\; x \ge 0)
\;\lor\;
(x \ge 6000 \;\land\; y = -x + 10000)
\;\lor\;
(x \ge 5000 \;\land\; x < 6000 \;\land\; y = -4x + 28000).
\)

Comparing the initial and final invariants, the phase boundaries (e.g., $x < 4000$, $4000 \le x < 5000$) are correctly identified from the outset, while only the arithmetic expressions within each phase require correction (e.g., $y = -3x + 25000$ is refined to $y = -4x + 28000$). This suggests that LLMs are effective at capturing the structural decomposition of multiphase programs, while precise numerical reasoning remains the primary source of error during generation.

This separation between structural and arithmetic errors explains why the framework remains robust even when the initial candidate is incorrect: because each refinement iteration targets a single \myterm{}'s arithmetic expression, errors in one branch do not propagate to others. We further observe that removing the localization mechanism causes this example to become unsolvable, providing direct evidence that localized refinement is essential for preventing cross-branch regression. Experimental evidence supporting these observations is presented in \autoref{subsec:RQ3}.

\subsubsection{Periodic Programs}

\input{figure/periodic-case-study}

\autoref{lst:example-unsolved} presents a program that goes beyond the multiphase setting targeted by \thistool{}. Unlike the monotonically progressing phases in the previous example, the modulo operation on $x$ introduces periodic behavior: execution repeatedly alternates between $x < 500$ and $x \ge 500$ with a fixed period of 1000, rather than progressing through separable phases.

In this setting, the relationship between $y$ and $z$ depends on how many times the condition $x < 500$ holds within each cycle of length 1000. Capturing this relationship requires reasoning over periodic dependencies and global iteration counts, which lies outside the scope of phase-local analysis and remains an open challenge for all evaluated tools.

A correct loop invariant does exist, as verified by an SMT solver. For example:
\(
(0 \le x \land x \le 500 \land z = y + x)
\;\lor\;
(x \ge 500 \land y < 1000 \land z = y + 1000 - x).
\)
Notably, this invariant still exhibits a disjunctive structure over branching conditions, suggesting that the decomposition strategy of \thistool{} provides a useful structural foundation. The remaining gap lies in reasoning about periodic phase transitions, which we consider a promising direction for future extension.

%% file: table/multiphase-results.tex
\begin{table}[t]
\centering
\small
\setlength{\tabcolsep}{4pt}
\caption{Results on the \textbf{multiphase benchmark}.}
\label{tab:multiphase-results}
\begin{tabular}{@{}lccc@{}}
\hline
\textbf{Method} & \textbf{\# Solved} & \textbf{Avg. Time (s)} & \textbf{Avg. Cost (\$)} \\
\hline
Clause2Inv & 20 & 51.95 & \textbf{0.018} \\
LaM4Inv & 25 & 112.69 & 0.709 \\
LORIS   & 18 & 200.07 & 0.223 \\
PAL2Inv & 29 & 104.014 & 0.174 \\
Code2Inv & 5 & 58.27 & N/A \\
LIPuS & 0 & 0.00 & N/A \\
CLN2INV & 1 & \textbf{1.1} & N/A \\
Eldarica & 20 & 94.76 & N/A \\
CPAchecker & 4 & 3.60 & N/A \\
UAutomizer & 19 & 23.05 & N/A \\
ImplCheck & 22 & 8.95 & N/A \\
\hline
\textbf{\thistool} & \textbf{35} & 167.48 & 0.226 \\
\hline
\end{tabular}
\end{table}

%% file: table/linear-nonlinear-results.tex
\begin{table}[t]
\centering
\small
\setlength{\tabcolsep}{4pt}
\caption{Results on Linear and Nonlinear(NL) benchmark.}
\label{tab:existing-results}
\begin{tabular}{@{}lcc@{}}
\hline
\textbf{Method} & \textbf{\# Solved (Linear / NL)} & \textbf{Avg. Time (s)} \\
\hline
Clause2Inv & 345 (307 / 38) & 14.41 (12.68 / 28.45) \\
LaM4Inv & 342 (\textbf{311} / 31) & 27.64 (23.66 / 67.58) \\
LORIS & 330 (287 / 43) & 46.83 (46.16 / 51.32) \\
PAL2Inv & 242 (227 / 15) & 37.78 (34.15 / 77.48) \\
Code2Inv & 205 (205 / 0) & 34.54 (34.54 / 0.00) \\
LIPuS & 159 (159 / 0) & 48.4 (48.4 / 0.00) \\
CLN2INV & 211 (211 / 0) & \textbf{0.6} (\textbf{0.6} / 0.00) \\
Eldarica & 256 (245 / 11) & 2.78 (2.02 / 19.52) \\
CPAchecker & 145(134 / 11) & 7.27(3.61 / 51.85) \\
UAutomizer & 321(292 / 29) & 13.00(11.09 / 32.17) \\
ImplCheck & 289 (284 / 5) & 2.02 (2.00 / \textbf{3.15}) \\
\hline
\textbf{\thistool} & \textbf{355 (311 / 44)} & 61.58 (51.83 / 130.51) \\
\hline
\end{tabular}
\end{table}

%% file: table/ablation-results.tex
\begin{table}[t]
\centering
\small
\setlength{\tabcolsep}{4pt}
\caption{Ablation study on the multiphase benchmark. \thistool-NL: No-Localization, and \thistool-CoT: CoT-Generation.}
\label{tab:ablation-results}
\begin{tabular}{@{}lccc@{}}
\hline
\textbf{Variant} & \textbf{\# Solved} & \textbf{Avg. Time (s)} & \textbf{Avg. Cost (\$)} \\
\hline
\textbf{\thistool{}} & \textbf{35} & 167.48 & 0.226 \\
\thistool-NL  & 27 & 134.25 & 0.197 \\
\thistool-CoT & 20 & \textbf{103.93} &  \textbf{0.103} \\
\hline
\end{tabular}
\end{table}

%% file: table/branch_count_result.tex
\begin{table}[t]
\centering
\small
\setlength{\tabcolsep}{5pt}
\caption{Performance of \thistool{} stratified by the number of non-exit loop-body branches.}
\label{tab:branch-count-results}
\begin{tabular}{@{}ccccc@{}}
\hline
\textbf{\# Branches} & \textbf{Benchmarks} & \textbf{Solved} & \textbf{Rate} & \textbf{\shortstack{Median\\Time (s)}} \\
\hline
1        & 176 & 174 & 98.9\% & 19.3 \\
2        & 139 & 119 & 85.6\% & 47.7 \\
3        & 33  & 31  & 93.9\% & 50.4 \\
4        & 55  & 50  & 90.9\% & 70.2 \\
$\geq 5$ & 17  & 16  & 94.1\% & 117.9 \\
\hline
\end{tabular}
\end{table}


%% file: figure/multiphase-case-study.tex
\begin{figure}[t]
\begin{lstlisting}[
    moredelim={**[is][\colorbox{yellow!25}]{@h@}{@/h@}},
]
int main() {
    int x, y;

    x = 0;  y = 0;          // pre-conditions

    while (unknown()) {      // loop body
@h@        if (x < 5000) {@/h@
@h@            if (x >= 4000) {@/h@
@h@                y = y + 4;@/h@
@h@            } else {@/h@
@h@                y = y + 1;@/h@
@h@            }@/h@
@h@        } else {@/h@
@h@            if (x >= 6000) {@/h@
@h@                y = y - 1;@/h@
@h@            } else {@/h@
@h@                y = y - 4;@/h@
@h@            }@/h@
@h@        }@/h@
@h@        x = x + 1;@/h@
    }

    if (x == 10000) {        // post-condition
        assert(y == 0);
    }
}
\end{lstlisting}
\caption{A representative multiphase loop with nested branches.}
\label{lst:example-multi}
\end{figure}

%% file: figure/periodic-case-study.tex
\begin{figure}[t]
\begin{lstlisting}[
    moredelim={**[is][\colorbox{yellow!25}]{@h@}{@/h@}},
]
int main() {
    int x, y, z;

    x = 0;  y = 0;  z = 0;   // pre-conditions

    while (unknown()) {       // loop body
@h@        if (x < 500) {@/h@
@h@            z = z + 2;@/h@
@h@        }@/h@
@h@        x = (x + 1) % 1000;@/h@
@h@        y = y + 1;@/h@
    }

    if (x == 0) {             // post-condition
        assert(y == z);
    }
}
\end{lstlisting}
\caption{Periodic multiphase loop with modulo.}
\label{lst:example-unsolved}
\end{figure}

%% file: content/6.Discussion.tex
\section{Discussion\label{sec:discuss}}

\noindent\textbf{Invariant synthesis vs.\ unreachability-based verification.}
Software model checkers such as CPAchecker~\cite{beyer2011cpachecker} and UAutomizer~\cite{heizmann2013ultimate} verify safety properties via unreachability checking without necessarily producing explicit inductive invariants. \thistool{} is not intended to replace such tools but targets the complementary problem of explicit invariant synthesis, which is needed in deductive verification~\cite{hoare1969axiomatic,leino2010dafny}, compositional reasoning with module-level contracts~\cite{calcagno2009bi,calcagno2015moving}, proof certificate generation~\cite{beyer2016reliable}, and neurosymbolic verification pipelines~\cite{wu2023lemur}. A systematic comparison of the two paradigms on diverse program classes is a direction for future work.

\noindent\textbf{Relationship to \textsc{PAL2Inv}.}
\textsc{PAL2Inv}~\cite{pal2inv} is the closest architectural baseline because both approaches expose the structure of \myterms{} during synthesis. It obtains initial invariants through abstract interpretation, generates atomic assertions for all enumerated \myterms{} in one LLM response, and uses counterexample-guided clause combination to construct disjunctive invariants from clause pools indexed by \myterm{}. \thistool{} instead queries the LLM separately for each \myterm{}, retains several complete candidate sub-invariants per component, and constructs a global candidate by selecting one candidate for each component. A concrete SMT counterexample identifies the component to replace while the other selections remain fixed, whereas \textsc{PAL2Inv} enriches and recombines its clause pools. The distinction concerns how the structure of \myterms{} supports independent generation and targeted replacement throughout the search. \textsc{PAL2Inv} solves 29 of 54 multiphase tasks compared with 35 for \thistool{}, with lower mean runtime and API cost (\autoref{tab:multiphase-results}). On the linear and nonlinear suites, \thistool{} solves 311 and 44 tasks, respectively, compared with 227 and 15 for \textsc{PAL2Inv} (\autoref{tab:existing-results}). The main reasons for this coverage advantage are as follows. First, PAL2Inv's abstract interpretation faces a precision--scalability trade-off: polyhedral analysis can become expensive as variables and constraints grow, while coarser domains may lose nonlinear or disjunctive relations. \thistool{} generates local candidates directly, avoiding this preprocessing cost and dependence on abstraction precision. Second, growing pools of atomic assertions still require the correct grouping into conjunctions and disjunctions, increasing the cost of constructing candidates and verifying them with SMT. \thistool{} generates complete local formulas, bounds retained global compositions through beam search, and repairs one selected component while preserving the others, reducing repeated reconstruction of useful combinations.

\noindent\textbf{Periodic control-flow patterns.}
As discussed in \autoref{sec:case_study}, \thistool{} has difficulty handling programs where execution repeatedly alternates between \myterms{} rather than progressing through separable phases. This limitation is shared by existing invariant synthesis techniques.

\noindent\textbf{Limits of heuristic recovery.}
When \textsc{IsAmbiguous} detects a counterexample crossing two \myterms{}, \textsc{JudgeBranch} uses an LLM to estimate whether the predecessor state is reachable and select the component to refine. This judgment is heuristic. Misclassifying a reachable state as unreachable can direct repair toward the source component when the destination component needs correction, and the reverse error can leave a spurious predecessor unexcluded. Such misjudgments can waste refinement rounds, discard useful local candidates, or prevent synthesis before the timeout. LLM stochasticity can also provide an opportunity for recovery: if a later iteration revisits an ambiguous failure, a new sample may identify the correct \myterm{} even when an earlier sample selected the wrong one. Candidate diversity and \textsc{SelectBest} provide further opportunities to explore alternative compositions. However, successive samples may repeat the same error, and the verifier rejects invalid candidates without certifying the reachability judgment itself. Recovery is therefore possible but not guaranteed. These errors affect search efficiency and success within the time budget, while soundness is preserved because every returned invariant must satisfy all SMT verification conditions.

\noindent\textbf{LLM stochasticity and robustness.}
Our design mitigates LLM stochasticity at two levels: localized refinement confines each edit to one sub-invariant, while beam search provides candidate diversity. The independent ablations (\autoref{subsec:RQ3}) support both contributions: removing localization reduces solved benchmarks from 35 to 27, while restricting search to one candidate per \myterm{} (CoT, $k{=}1$) while retaining localization reduces coverage from 35 to 20.

\noindent\textbf{Choice of LLM.}
All LLM-based tools use the same model (GPT-5.1) to ensure a fair comparison. The framework is model-agnostic in that it relies only on standard text generation with structured JSON output and does not depend on any model-specific features. Using one model isolates the contribution of our framework architecture from confounds introduced by varying model capabilities.


\noindent\textbf{Data contamination.}
Due to the opaque nature of LLM training data, benchmark programs may have been seen during pretraining. Our ablation study (\autoref{subsec:RQ3}) suggests the model is not simply memorizing solutions, but this threat applies to all LLM-based baselines and cannot be entirely eliminated.

%% file: content/7.RelatedWork.tex
\section{Related Work \label{sec:related_work}}


\noindent\textbf{Traditional and learning-based loop invariant synthesis.}
Traditional approaches rely on symbolic techniques such as abstract interpretation~\cite{cousot1977abstract,cousot1979systematic,karr1976affine, cousot1978automatic}, abduction~\cite{calcagno2009bi,dillig2013inductive}, dynamic analysis~\cite{ernst2007daikon,echenim2019ilinva,lattner2008llvm}, model checking~\cite{hojjat2018eldarica,vediramana2024global}, Craig interpolation~\cite{jhala2006practical,mcmillan2010lazy}, constraint-based and SyGuS methods~\cite{colon2003linear,gupta2009tests,barbosa2022cvc5,barrett2011cvc4}, and CEGAR~\cite{clarke2003counterexample,zhu2015learning}.
On the learning side, the ICE framework~\cite{garg2014ice} introduced a foundational paradigm of learning from examples, counterexamples, and implications, inspiring decision-tree-based~\cite{garg2016learning,ezudheen2018horn,riley2022multi,xu2020interval}, classification-based~\cite{Lu2020PBLInv:}, and reinforcement-learning-guided~\cite{Liu2023Towards} methods.
Representative learning-based tools include Code2Inv~\cite{si2020code2inv}, which learns candidates with neural models, CLN2INV~\cite{ryan2019cln2inv}, which uses continuous logic networks, and LIPuS~\cite{yu2023lipus}, which combines reinforcement learning with SMT solving.
ImplCheck~\cite{riley2022multi} uses model-based projection to discover phase guards and data learning to synthesize implication invariants, including for mutually dependent periodic phases.
AutoINF~\cite{hadad2026autoinf} models loop paths with a path dependency automaton and uses targeted counterexamples to refine polynomial invariant templates. Its trace-based template inference offers a symbolic comparison with our LLM-generated local candidates and search over complete compositions.
These approaches differ in their supported invariant representations, abstractions, and search strategies.
Our work leverages the generative capacity of LLMs and specifically targets multiphase programs where invariants must capture phase-specific behaviors.
We note that software model checkers such as CPAchecker~\cite{beyer2011cpachecker} and UAutomizer~\cite{heizmann2013ultimate} address the orthogonal problem of unreachability-based verification without necessarily producing explicit invariants, and a detailed discussion of the complementary roles of invariant synthesis and unreachability checking is provided in \autoref{sec:discuss}.

\noindent\textbf{LLM-based loop invariant synthesis.}
Recent work leverages large language models (LLMs) to generate loop invariants, each adopting a different decomposition strategy.
ACInv~\cite{LIU2026103387} targets data-structure-rich programs by combining static analysis with an LLM-based evaluator for iterative refinement.
\toolb{}~\cite{cao2025clause2inv} decomposes at the clause level, generating individual clauses that are assembled and validated by an SMT solver.
\toola{}~\cite{wu2024llm} filters LLM-generated predicates via bounded model checking before reassembling them for Z3 verification.
LORIS~\cite{li2026loris} improves feedback quality by translating the LLM's reasoning steps into first-order logic implications to pinpoint logical flaws.
PAL2Inv~\cite{pal2inv} combines abstract interpretation with atomic assertions generated for enumerated loop branch sequences and counterexample-guided clause combination. It maintains pools of assertions for these sequences, whereas \thistool{} separately generates and retains complete local candidate formulas for targeted replacement.
Our work differs in two fundamental ways: compositional inference decomposes invariant synthesis along \myterm{}s so that the LLM reasons about one execution scenario at a time, and counterexample-guided localized refinement traces verification failures to the specific failing \myterm{} and repairs only that phase's sub-invariant. This phase-level repair capability is unavailable to whole-loop methods.

\noindent\textbf{LLM-based formal verification.}
LLMs have also been applied to broader formal verification tasks.
Baldur~\cite{first2023baldur} demonstrates that LLMs can generate and repair complete proofs, while Lemur~\cite{wu2023lemur} provides a formally sound framework for integrating LLMs with SMT solvers.
Cai et al.~\cite{cai2024towards,cai2025automated} apply refinement calculus to guide LLM-based program synthesis with formal guarantees, and Wen et al.~\cite{wen2024enchanting} combine LLMs with static analysis to synthesize full program specifications.
Other studies evaluate LLM-assisted verification workflows~\cite{kamath2024leveraging,wang2025supporting,couder2024requirements} or survey LLM trustworthiness from a verification perspective~\cite{huang2024survey}.
These works establish the viability of LLM-based formal reasoning and inform the design of our framework, which applies LLM--solver integration specifically to the problem of multiphase invariant synthesis.

\noindent\textbf{Structured reasoning with LLMs.}
Tree-of-Thought~\cite{yao2023tree,long2023large} and self-consistency~\cite{wang2022self} extend chain-of-thought prompting~\cite{wei2022chain} by exploring multiple reasoning paths via tree search or sampling. These methods use generic decompositions with LLM self-evaluation at each node.
\thistool{} differs in two key respects: its tree topology is derived from the program's branch semantics (compositional inference) rather than generic reasoning steps, and refinement is guided by SMT counterexamples that precisely localize the failing \myterm{} (localized refinement) rather than relying on LLM self-evaluation for backtracking decisions.

%% file: content/8.Conclusion.tex
\section{Conclusion} \label{sec:conclusion}

We present \thistool{}, a framework for loop invariant inference on multiphase programs. Compositional inference decomposes synthesis along \myterm{}s so that the LLM reasons about one execution scenario at a time, while counterexample-guided localized refinement traces failures to the responsible \myterm{} and repairs only that sub-invariant. Experiments on 420 benchmarks show that \thistool{} achieves state-of-the-art performance on multiphase programs and remains competitive on linear/non-linear benchmarks.

